\documentclass[journal]{IEEEtran}
\usepackage{cite}
\usepackage{graphicx}
\usepackage{framed}
\usepackage{blindtext}
\usepackage{comment}
\usepackage{color}
\graphicspath{ {images/} }

\makeatletter
\def\ps@IEEEtitlepagestyle{%
  \def\@oddfoot{\mycopyrightnotice}%
  \def\@evenfoot{}%
}
\def\mycopyrightnotice{%
  \gdef\mycopyrightnotice{}
}

\begin{document}

\title{
Block4Forensic: An Integrated Lightweight Blockchain Framework for Forensics Applications of Connected Vehicles
}
\author
{\IEEEauthorblockN{Mumin Cebe, Enes Erdin, Kemal Akkaya, Hidayet Aksu and Selcuk Uluagac}
\IEEEauthorblockA{Department of Electrical and Computer Engineering\\
Florida International University\\
Miami, FL 33174\\
Email: \{mcebe|eerdi001|kakkaya|haksu|suluagac\}@fiu.edu}
}
\maketitle

\begin{abstract}
Today's vehicles are becoming cyber-physical systems that do not only communicate with other vehicles but also gather various information from hundreds of sensors within them. These developments help create smart and connected (e.g., self-driving) vehicles that will introduce significant information to drivers, manufactures, insurance companies and maintenance service providers for various applications. One such application that is becoming crucial with the introduction of self-driving cars is the forensic analysis for traffic accidents. The utilization of vehicle related data can be instrumental in post-accident scenarios to find out the faulty party, particularly for self-driving vehicles. With the opportunity of being able to access various information on the cars, we propose a permissioned blockchain framework among the various elements involved to manage the collected vehicle related data. Specifically, we first integrate Vehicular Public Key Management (VPKI) to the proposed blockchain to provide membership establishment and privacy. Next, we design a fragmented ledger that will store detailed data related to vehicle such as maintenance information/history, car diagnosis reports, etc. The proposed forensic framework enables trustless, traceable and privacy-aware post-accident analysis with minimal storage and processing overhead.

\end{abstract}

\section{Introduction}
\label{sec:intro}
Today's vehicles are becoming much smarter with special-purpose sensors, control units and wireless adapters to monitor their operations and communicate with their surroundings \cite{wave2016}. These contemporary \emph{smart vehicles} are now considered as a comprehensive cyber-physical systems (CPS) with communication, control and sensing components \cite{berger2014autonomous}. For instance, Electronic Control Units (ECU) and On Board Units (OBUs) can receive data from various on-board sensing devices to take certain actions. The connections among the control units and sensor devices are done via different types of networks such as  
Controller Area Network (CAN) Bus, 
Local Interconnect Network (LIN)
Bus, Flexray, Bluetooth, etc. Such developments along with capabilities to sense and communicate with the surroundings are enabling further developments such as creation of autonomous vehicles which are also known as self-driving cars 
that will revolutionize our lives.  


The penetration of IoT technologies in vehicles enables collection of enormous data from vehicles for various applications. For instance, most vehicles that are manufactured in the last decade have on-board diagnostics (OBD) ports which are used for retrieving vehicle controller diagnostics. These ports are typically interfaced with a Wi-Fi, Bluetooth or serial connection to supply data outside.
Another major development is the deployment of Event Data Recorders (EDRs) by leading manufacturers such as GM, Ford, etc. 
EDRs are meant to store incident data based on triggering events. Finally, the future vehicles will be equipped with On-board Units (OBUs) to enable connectivity among vehicles and Road Side Units (RSUs) to provide collision avoidance and congestion control. Such safety features will be realized with the wireless Dedicated Short Range Communications (DSRC) that will not only enable broadcasting of basic safety messages (BSM) (i.e., Vehicle-to-Vehicle (V2V)) but also provide the means to communicate with the infrastructure such as the traffic lights, railroad crossing etc. (i.e., Vehicle-to-Infrastructure (V2I)). Although BSM is name of a spacial message in DSRC specification, here it is used as a generic name allocating all safety related messages \cite{wave2016}. 


Such capabilities as being able to collect data within and around the vehicles can make a significant impact on vehicular forensics that aims to investigate the reasons behind the accidents. This field will become even more important with the proliferation of self-driving cars which are prone to failures and cyber attacks \cite{baig2017future}. Typically, after an accident, investigator specialists 
analyze the causes of the accident so that disputes among parties can be resolved. The investigators look at many different aspects including inspection of accident site and vehicle. Site inspection contains physical evidences such as scrub marks, position of vehicles, tire conditions etc. 
In addition to those physical evidences digital data supplied from OBD ports and EDR introduce valuable complementary evidence for supporting the dispute resolution.
Eventually, by enabling to capture, store and transfer the vehicle  data, the puzzle including drivers, insurance companies, manufacturers and law enforcement authorities can be solved \cite{baig2017future,mansor2016log}.


Even after utilizing EDR and OBD data, the accident investigation lacks certain features which are absolutely needed for a comprehensive dispute resolution. These can be listed as follows: 1) The obtained data does not include a comprehensive history of the vehicle due to limited storage 
(i.e., the data is overwritten after a while); 2) The parties do not have direct control on the extracted data, therefore they should trust third parties 
which incurs questions about the integrity of data
3) There is no system for integrating data from all parties including other vehicles, road conditions, manufacturers and maintenance centers; 4) There is no vehicular forensics solution to resolve a hit and run case other than third party information such as surveillance cameras and eyewitnesses.

Therefore, in this paper, we address these points by proposing Block4Forensic (B4F) framework, a blockchain-based vehicular forensics system that will collect vehicles and related business components under the same umbrella. In particular, the proposed system 1) provides a lightweight  privacy-aware blockchain by gathering all related parties such as drivers, maintenance centers, car manufacturers and law enforcement without requiring a trusted third party in case of an incident; and 2) introduces a vehicular forensics investigation framework that harbors all necessary data for a comprehensive vehicular forensics solution.


The rest of the paper is organized as follows. In Section II, we describe the preliminaries related to all concepts and provide a summary of the state of the art. In Section III we introduce B4F framework. In Section IV, we explain BF4 with its components. Section V is dedicated to future issues in this emerging research area. Finally, we conclude the paper in Section VI.  

\section{Background}

\subsubsection{Vehicular Forensics}
Traditional vehicular forensics deals with the physical evidences collected from an accident scene, such as photographs, measurements, scrub marks etc.
Usage of vehicle generated data has attracted the interest of the researchers ~\cite{nilsson2008conducting}, hence it is strengthening the hands of the forensic investigators as they can find supporting evidences from the digital subsystems of a vehicle. There are many controllers and sensors in modern vehicles in different capabilities. For a better driving experience almost every capability of the vehicle is measured and reported. 



\begin{figure}
	\centering
	\includegraphics[width=0.65\linewidth]{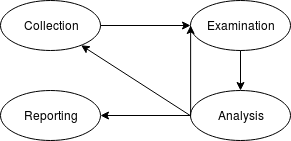}
	\caption{Digital Forensics Process Model \cite{karabiyik2015building}}
	\label{process}
\end{figure}

When an accident occurs, incident first responders arrive to the scene to identify and secure the digital devices for keeping them forensically sound (preserving integrity of evidence) by following the process shown in shown in Fig. \ref{process}. After securing  and getting access to all related devices, further examination and analysis are performed. This basically means finding incident related data on the digital devices such as finding traces of a cyber attack and failure of a manufacturer component or mistake of a driver etc.
At the reporting phase, investigators prepare a report and 
testify and present the evidence.
Obviously, the most important factor for the admissibility of the report is to verify that the evidence devices have not been altered during the investigation. 
This may be quite challenging as there is no universal standard to collect, examine and analyze data from digital devices on vehicles, drivers, 
and involved units.
Therefore, a framework that will enable convenient data collection and analysis is needed. The framework should satisfy privacy of the user 
and the stored data content should be clear to the user i.e., owner of the data.

\subsubsection{Event Data Recorders and On Board Diagnosis }
Event data recorder (EDR), informally named as ``black box'' is a device placed in vehicles in order to collect data related to crashes and accidents. In case of a dispute, investigators come up with the most probable set up. The digital data recorded by the EDR is widely used as a supporting evidence in investigations for reconstructing the accident scene. When a triggering event occurs - some among those events are the airbag deployment, sudden speed changes above a threshold - EDR captures and stores the state of the vehicle in a tamper proof storage. It is known that EDR data is extracted by the investigators through the on board diagnosis (OBD) port in case of an incident.  Meanwhile, the ownership of EDR data and integrity of it is discussed in study \cite{kowalick500motor} along with how this data is used by Traffic Safety Administrator (NHTSA) and other third parties for post-accident scenario reconstruction.


\subsubsection{Dedicated Short Range Communications (DSRC) and Basic Safety Messages (BSM)}
DSRC specification defines the dedicated channels, standards and protocols for communication between connected vehicles.  
Among many different messages, BSM is one of the most important one for safety-related awareness between the vehicles. 
Part I of the BSM includes high priority information about a vehicle such as position, speed, size, brake status and ID of the vehicle and also medium priority messages such as positional accuracy and steering wheel angle. 
This scheme brings additional value in the forensic investigation since the collected digital data will not be solely related to the car itself but also related to the participants surrounding it. 

\subsection{Vehicular Public-Key Infrastructure - Vehicular Network Security}
In the networking layer of communication of connected vehicles, IEEE 1609.2 standard is utilized for message integrity and authentication \cite{IEEE2016_2}.

\begin{figure}
\centering
\includegraphics[width=\linewidth]{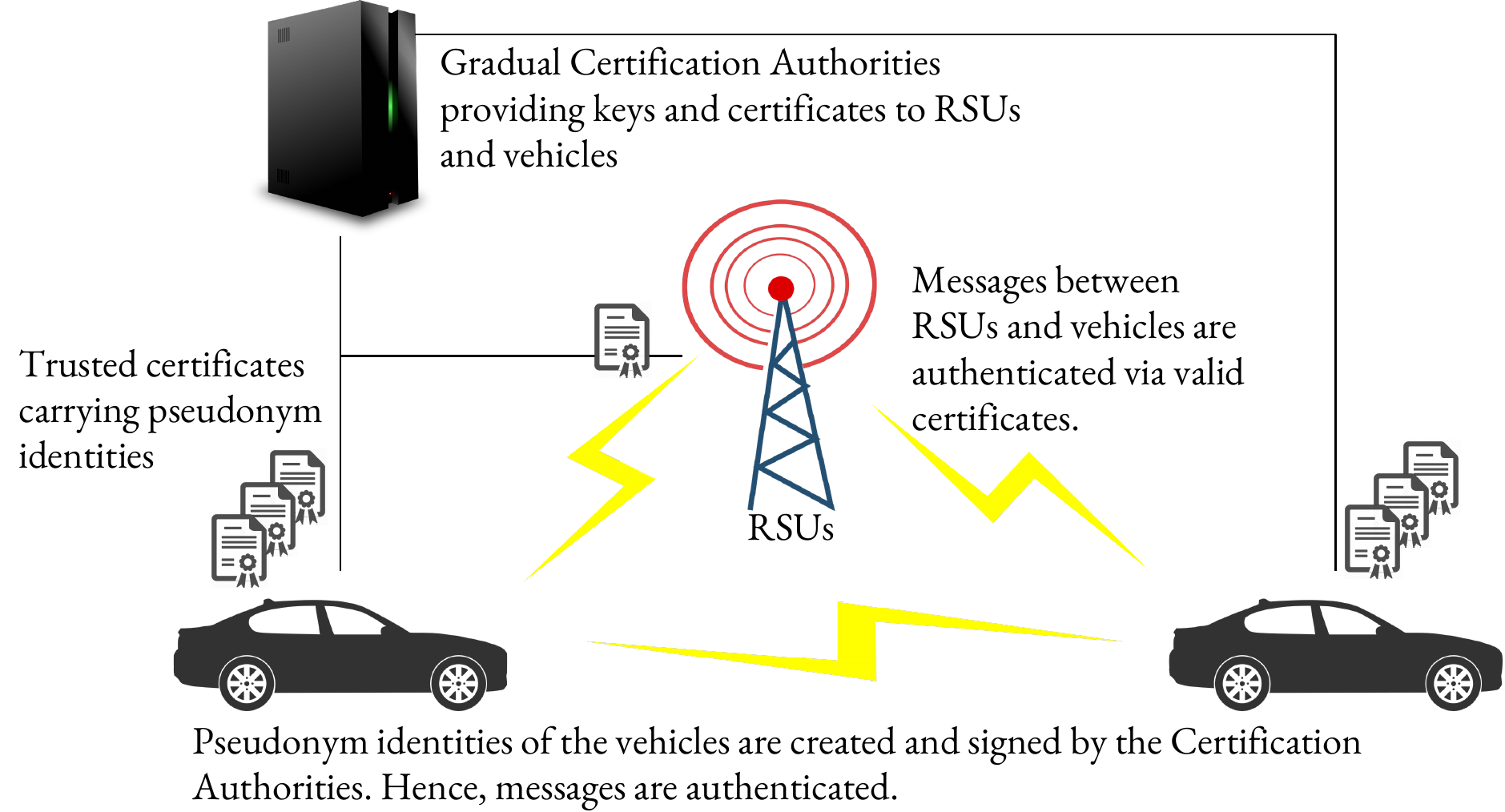}
\caption{\small A simplified representation of VPKI.}\label{fig:vpki}
\end{figure}

The vehicular public-key infrastructure (VPKI), a simplified version of which is shown in Fig. \ref{fig:vpki}, utilized in IEEE 1609.2 is a highly complicated infrastructure specially tailored for the needs of the transportation system. The main certification authority (CA) 
generates, distributes, and revokes the digital certificates. Proposed VPKI structure also deals with the privacy and security issues. According to the safety pilot model, the certificates which constitute the pseudonym identity of the vehicle are valid for only 5 minutes. That behavior provides anonymity for the communicating parties and also it makes the system strong against targeted attacks aiming privacy and spoofing. 

\subsection{Blockchain}
A blockchain is composed of blocks which are linked to each other and secured cryptographically. 
This establishes a strong tie between blocks which guarantees the  order of blocks and provides an implicit strong timestamp mechanism, thus a block is prevented from any altering without changing all of its successors. This blockchain data structure can be shared 
to build a distributed data structure called as \emph{shared ledger} \cite{nakamoto2008bitcoin}.
This working scheme of blockchain carries unique properties such as relieving central authority trust, immutability, and timestamping.



There are two types of blockchain structure, public and permissioned. 
For instance,
Bitcoin 
and Ethereum 
fall into public blockchain category where everyone is able to read and write the ledger without any restriction (i.e., there is no membership requirement). However, in permissioned blockchains \cite{cachin2016architecture}, the participants forms a members only club.

The process of adding a new block to the chain is carried out via a protocol, which establishes consensus among participants to confirm the new block.
The implementation details of consensus protocol (\textcolor{red}{e.g.}, proof-of-work or POW) changes a lot depending on the type of blockchain 
For instance, in public blockchain, consensus is typically  a form of hash puzzle which requires finding a predefined hash value.
 This consensus protocol brings a significant level of security on the chain (withstand up to 50\% of nodes are being malicious), but at the cost of computational power and time. For instance, Bitcoin's maximum throughput is  7 transactions per second and the consensus finality can take an hour.
On the other hand, permissioned blockchains utilize some kind of Byzantine fault tolerant voting based algorithm as consensus mechanism, such as Practical Byzantine Fault Tolerance (PBFT) 
or Stellar Consensus Protocol (SCP)
, which do not require computationally expensive hash puzzle. 
As a result, reaching a consensus is faster which means higher transaction throughput. However, permissioned blockchains generally require more than two-thirds of nodes to be trustworthy rather than 51\%.
More details about consensus algorithms can be found in \cite{bano2017consensus}.

\subsection{Current State of the Art in Vehicular Forensics}
The use of the Digital Vehicular Forensics is increasing in investigations. There are commercial products targeting comprehensive data collection from the cars. iVe project from Berla is a result of that effort, where their product has access to EDR and OBD port. They also retrieve data from the infotainment and telematics systems. The data is collected on cloud storage. Authors in \cite{mansor2016log} offer a similar solution. EDR and OBD ports are accessible by design and the data is stored on the cloud. Although the authors in \cite{ab2016forensic} do not directly aim implementation for digital forensics, they offer a framework mainly discussing guidelines named ``forensic by design''. The idea of blockchain utilization for vehicular security is offered in \cite{dorri2017blockchain}. The authors sketch possible use cases for insurance companies or wireless software updates for smart cars, however, their discussion lacks practical issues such as membership management and scalability. For a proper investigation non-repudiation is of great importance. There is an implicit consensus in the research community that public key cryptography produces reliable solutions for that issue \cite{li2015acpn}. However, there is a need for a comprehensive applicable and scalable framework in vehicular forensics research.

\section{A Blockchain Framework Vehicular Forensic}
The ultimate aim of vehicular forensics is to resolve disputes and determine the faulty parts in case of an accident. The developments in \emph{connected vehicles} provide new opportunities for forensic analysis by taking advantage of the IoT and CPS features. 
Utilizing produced sensors data with decision entities would allow building a comprehensive vehicular forensic analysis. Considering involving multiple parties such as manufacturers, drivers, insurance companies, law enforcement etc., we first identify the key features for an effective and trustworthy vehicular forensics framework. 

\subsection{Desired Features of Envisioned Forensic Analysis}
\label{sec:reqs}
The following key features are desired for Vehicular Forensics: 

\begin{itemize}
\item{\textbf{Integrity:}} The integrity of forensic data is very important for resolving the disputes.
\item{\textbf{Non-Repudiation:}} The parties should be held responsible for their actions by providing proof of the integrity. 
\item{\textbf{Relieve Single Point-of-Trust:}} The system should remove the assumption of trust reliance solely on a single authority and provide an accountable trustworthiness of each participant. 
\item{\textbf{Comprehensive Forensic Analysis:}} The system should provide a comprehensive mechanism to accident analysis by providing access to historical data even before the accident. For example, the behavioral pattern of vehicle after maintenance (e.g., steering ability, braking distance) or a previously reported malfunctioning component of a vehicle can provide important clues to determine the faulty party.
\item{\textbf{Lightweightness:}} The system should have minimum overhead on endpoints since it includes multiple parties that may have different capabilities and resources.
\item{\textbf{Privacy:}} The system should preserve the privacy of the participants while also providing the flexibility for the participants to selectively reveal their data as they wish. 
\end{itemize}

\subsection{B4F Framework}

\begin{figure}
\centering
\includegraphics[width=\linewidth]{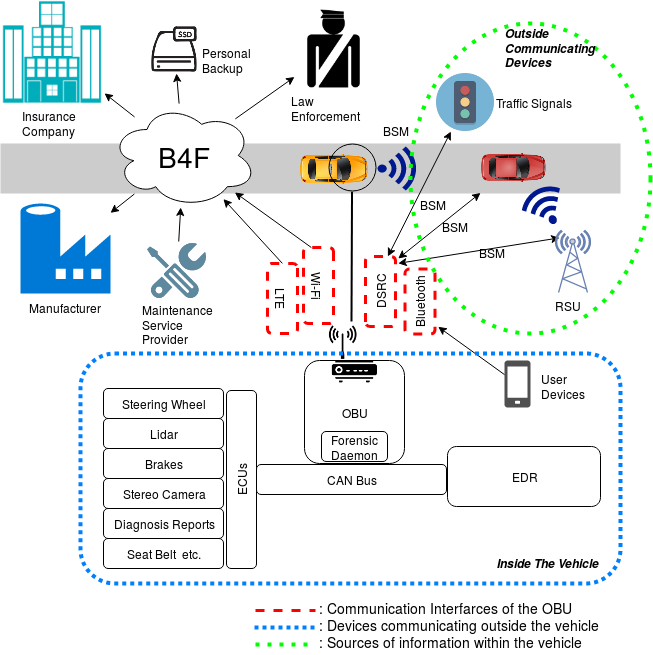}
\caption{\small An Overview of Forensic System Model with its Stakeholders.}\label{fig:overview}
\end{figure}

To enable the vehicular forensics vision, we introduce a novel blockhchain forensic framework shown in Fig.~\ref{fig:overview}. The framework connects the following stakeholders: vehicles, 
maintenance service providers (e.g., mechanics), vehicle manufacturers, law enforcement and insurance companies. 
The key features of the envisioned vehicular forensics system mentioned in the previous subsection guided us while building the blockchain-based vehicular forensics system. 

At the heart of design, there is a special \textit{forensic daemon}, which is stationed within the OBU and constantly retrieves data from EDR, BSMs (i.e., messages received from other vehicles) and on-board sensors/IoT devices through CAN Bus. The \textit{forensic daemon} periodically shares the EDR and BSM data with the insurance company 
through an encrypted channel.
Note that, only related BSMs are shared when an EDR triggering event occurs. On the other hand, the car manufacturers collect regular car diagnostic reports. A cryptographic hash of these data is submitted to Blockchain for removing single trust issue. Both insurance companies and manufacturers collect those data for analysis. 
Moreover, maintenance records are kept at the maintenance service providers, and a hash of each record is submitted to Blockchain in the same manner. As an optional extension to the framework all of mentioned data can also be stored on a personal cloud storage.

Stored data will be used in post-accident scenarios by allowing the parties to disclose their data selectively to determine the faulty party. Law enforcement authorities play an investigative role for post-accident scenarios while parties disclose their data with proof of integrity.


\subsection{Potential accident scene}

\begin{figure*}
\centering
\includegraphics[width=.75\linewidth]{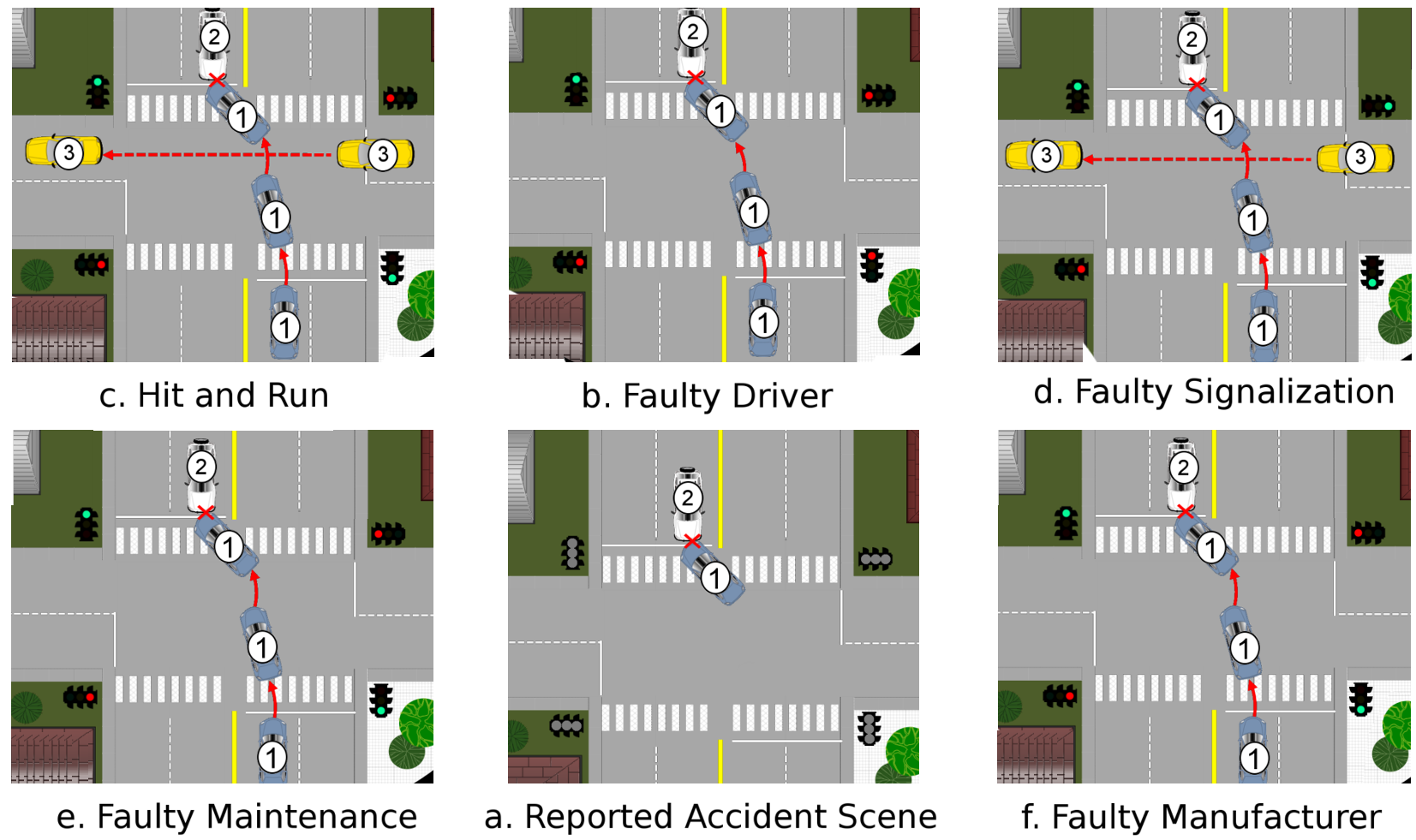}
\caption{\small A Hypothetical accident scene and possible reconstructions of the accident. }\label{fig:scenario}
\end{figure*} 

An investigator working on an accident scene needs to collect all pieces of clues to reconstruct the accident scene. Once the accident scene is reconstructed, the faulty party can be determined accordingly. 


Here, we discuss how digital data provided by B4F assist an investigation. Assume that an accident scene where Vehicle 1 (V1) collided with Vehicle  2 (V2) at an intersection with traffic lights as illustrated in Fig.~\ref{fig:scenario}~(a). The data provided by B4F may enable various forensically sound scene reconstructions as listed below:
\begin{itemize}
\item \textbf{Reconstructed Scene (b)}: BSM messages include the traffic light status and  cars' last positions. In this scenario, BSM messages reveal that V1 started to turn left when the red light was on as shown in Fig.~\ref{fig:scenario}~(b). Lights status 
are being disseminated by the smart traffic lights and thus when the accident happens, B4F would have stored the last BSM messages from the traffic lights. Here, data clearly points out that the V1 is the faulty party. 

\item \textbf{Reconstructed Scene (c)}: Timestamped  data in B4F reveals the existence of another vehicle in the accident scene. Drivers of V1 and V2 started crossing the road when the light turned to green. At that time V3  did not stop at red light and caused V2 lost its control to hit V1. B4F data uncovers the existence of V3 and resolves such a hit case where faulty party is a third car that run out of the incident area.

\item \textbf{Reconstructed Scene (d)}: Similar to Scene (c), data reveals the existence of V3. However, this time none of the cars violate the rules as the traffic light for V3 is also green. BSM data supplied from smart traffic lights 
would reveal faulty signalization as the cause of the accident. 

\item \textbf{Reconstructed Scene (e)}: In this scenario B4F data indicates that none of the drivers violates the traffic rules. However, by investigating the car diagnostic report history on B4F, the investigator finds out that after maintenance, the vehicle has pulling problem while braking. Due to this faulty operation in V1, the driver lost the control of the car and hit V2. The history of previous vehicle maintenance records helps to resolve this complicated scenario and suggests maintenance provider as the faulty party. 

\item \textbf{Reconstructed Scene (f)}: In this scenario, B4F data shows that V1 was on the autopilot at the accident time. Moreover, the diagnostic records report a failed sensor. Thus, V1 autopilot software with faulty input caused the accident which suggests car manufacturer as the faulty party. 

\end{itemize}

Various parties might be involved in an accident like exemplified above. Forensic data provided by B4F provides a fast and efficient accident scene reconstruction which helps any investigation significantly. 




\section {B4F Components}
In this section, we 
first describe the forensics elements and data types and then we move on to elaborate  on the specific elements of B4F that relates to the blockchain structure, its membership management, and storage issues.

\subsection{Forensic Daemon}

Here, we explain how the proposed \textit{forensic daemon} interacts with different components of a vehicle. 
Note that our \textit{forensic daemon} runs as an application in OBU
thanks to existing SDKs for custom application development.

The OBU has a read access to the vehicle network infrastructure. The backbone of the vehicle network is the CAN bus.
In a modern vehicle, many important sub-structures like steering wheel motor, braking system,  throttle, tire pressure monitoring system, seat belt buckle status even windshield wipers are controlled and monitored via CAN bus. Thus, the CAN bus may deliver invaluable data in terms of vehicular forensics to the OBU which can be retrieved by the \textit{forensic daemon}.


Additionally, through Wi-Fi or Bluetooth interfaces, the \textit{forensic daemon} can receive data from the driver about his/her health status via wearables. Similarly, road conditions and weather data can be retrieved from RSUs or driver's smart phone that has applications related to such data.  

The \textit{forensic daemon} will collect data on predefined occasions based on basic or custom rules.
After adding a timestamp, it will sign the data using pseudonym certificate which is readily available in the OBU. In case of an investigation, submitted data 
will be disclosed for investigation by the user.


\subsection{Forensic data types and B4F process }
In this sub-section, we detail the interaction between vehicle and the B4F framework. There are 3 types of data in our framework. The first one is event data which is occurred in case of an incident triggered by the predefined conditions in EDR. The second one is the diagnosis data which is produced by the vehicle periodically or in case of a failure. Finally, there is a maintenance data which contains information about the maintenance report and kept by both maintenance service and user. Maintenance data is signed by both the vehicle and the maintenance provider so is a multi-signature data.
We have two data submission processes in the B4F. As described in Section IV.E, the content of forensic data along with time and pseudonym vehicle ID is signed by vehicle and submitted to the corresponding parties such as insurance companies, manufacturers and personal cloud storage. While the content of the forensic data is kept between two parties, the hash of this data is stored in shared ledger on blockchain. B4F implements a gossip network where each vehicle selects a random set of validators to gossip hash of data. To ensure that messages are valid, every message is signed by the pseudonym identity of the vehicle; validators check that the signature is valid before relaying it. The randomly chosen leader propose a block in submitted transactions and 
distributes its block of pending transactions through the
gossip protocol again. B4F establishes a byzantine agreement to reach the finality.

\color{black}

\subsection{Blockchain Structure}

To address the requirements of Section \ref{sec:reqs}, we propose utilizing \emph{permissioned} blockchain technology and implement \textit{shared} and \textit{fragmented} ledgers to securely and efficiently exchange information between the collaborating parties. 

%

\begin{figure}
\centering
\includegraphics[width=0.7\linewidth]{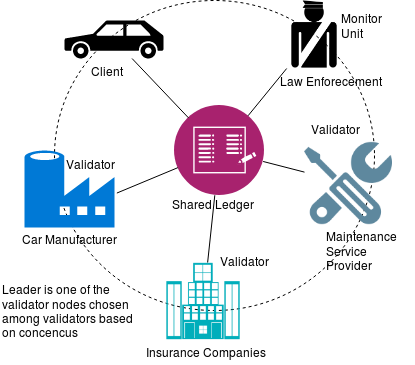}
\caption{\small Permissioned blockchain participants.}\label{fig:blockchain}
\end{figure}

In our proposed blockchain, we will have \textcolor{red}{four} different types of nodes: 1) Leader; 2) Validator; 3) Monitor units and 4) Client as shown in Fig.~\ref{fig:blockchain}. 


A leader is selected randomly every block time among the validator nodes (i.e. manufacturers, maintenance centers, insurance companies).
The client, i.e. vehicle, provides signed transactions to the B4F to ensure that messages cannot be forged.



The randomly chosen leader proposes a block to the network  based on the transactions it has received. To reach a consensus on a proposed block, validators run byzantine agreement protocols such as PBFT. These protocols are resilient to malicious actions of the leader and participants \cite{bano2017consensus}.
Monitor units are law enforcement authorities who do not directly participate in validation process, but keep  the replica of the shared ledger to be able to participate in post-accident disputes. 

This proposed framework is geared for increasing the level of trust among network participants and thus will eliminate the need to a trusted third party.

Due to use of hashes, the overhead of the building and storing replicated shared ledger among parties is minimized. Note that the integrity of data can be verified by comparing its hash value with the corresponding hashes that are stored on the Blockchain.

\subsection{Integrated Membership Management and Privacy via Pseudonym Certificates}

In a public blockchain anyone can participate either as a client or as a validator (e.g. miners in cryptocurrencies). 
However, in case of a permissioned blockchain, access permission is strictly controlled by membership service and only granted users are able to make transactions. The identities issued by membership service are unique and can not be altered. Thus, there is no support to protect privacy between interacting peers. Leveraging permissioned-blockchain impedes the use of anonymous identities contrary to identities used in public Blockchain such as Bitcoin. This is particularly important in our case, since the vehicle owners would like to protect their privacy while sharing data with their manufacturers and insurance companies. On the other hand, the huge number of network participants (e.g., millions of vehicles on the roads) expose membership management as a challenge in realization of permissioned blockchain.

Thus, we use  pseudonym identities from VPKI model suggested in IEEE 1609.2 as a token for clients to satisfy anonymity (i.e., vehicles) in the proposed B4F. 
According to the VPKI scheme, the vehicle has different pseudonym identities for different time intervals (i.e, every 5 minutes), thus every transaction will be submitted with different identity which protects the user privacy as defined in the attack model of IEEE 1609.2. However, regulations and policies should be assessed for a proper disclosure of the user data. In addition, exploiting VPKI scheme also addresses mentioned membership management challenge. Any vehicle that has a valid pseudonym identity can make transactions on the proposed Blockchain, since participants of B4F recognizes valid certificates produced by VPKI. Validator nodes check the validity of the certificate and timestamp of the submitted data (i.e., hash of the forensic data). 
If the timestamp belongs to the certificate validity period (i.e., every 5 minutes), the transaction is confirmed. 
The consensus on valid transactions is achieved by a  computationally inexpensive voting-based byzantine agreement scheme among validators.

\subsection{ Lightweight Fragmented Ledger for Forensic Participants}
\label{sec:fragmented}
Blockchain is a shared ledger that maintains a growing list of blocks that are chained to each other.  
Each participant stores a copy of the entire history. In our case, the data are immense and thus the shared ledger can grow dramatically and may cause both communication  and storage overhead. 

\begin{figure}
\centering
\includegraphics[width=\linewidth]{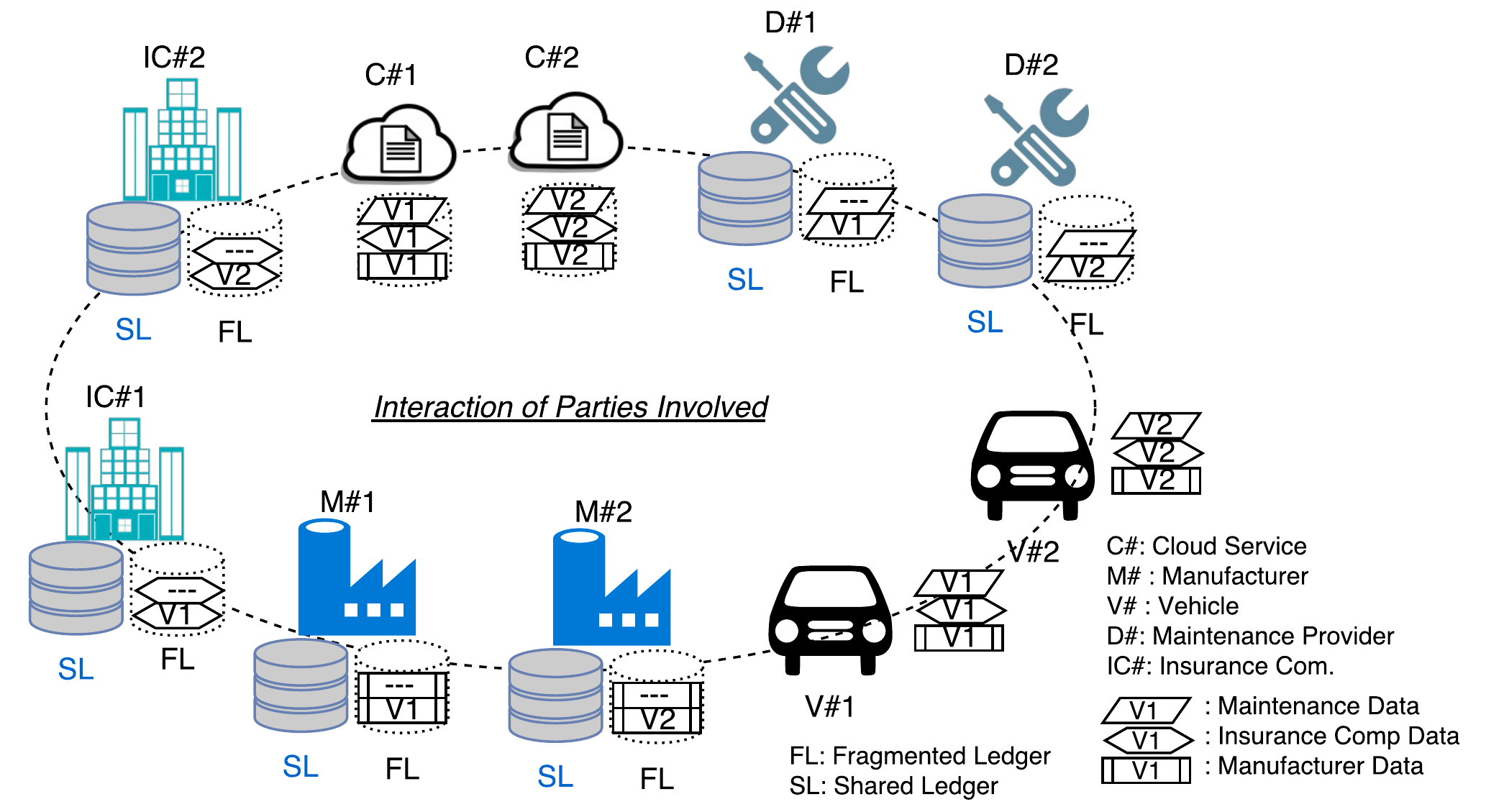}
\caption{\small An Overview of  Proposed Ledger Structure. SL represents shared ledger while FL represents fragmented ledgers which can hold different data. }\label{fig:fragmented}
\end{figure}


To address this issue, we utilize a \textit{fragmented ledger} instead of storing all forensic data in shared ledger.
The motivation comes from the observation that each party has already stored a different fragment of required data. For instance, a maintenance provider may not be interested in the content of periodic EDR data and thus there is no need to keep that content in shared ledger. On the contrary, as insurance companies keep EDR data in their fragmented ledger, keeping proof of that data in the shared ledger is sufficient.
Therefore, in B4F, all participants of the  network will have a consensus on the shared ledger. However, each participant maintains just related information which differs from others as shown in Fig.~\ref{fig:fragmented}. Specifically, the difference between the shared and fragmented ledgers will be on forensic data details. The shared ledger does not carry any information related to the forensic content of EDR\&BSM data, car diagnostic reports, provided maintenance etc. 

Additionally, note that the user may want to  refuse to submit maintenance or manufacturer data content. Instead, s/he keeps it in a personal cloud storage. However, based on regulations and policies, in case of an incident, the authorities will require the user to disclose this data as needed, integrity of which is satisfied by the Blockchain.

\color{black}

\section{Future Research Issues}
As there is a growing research on various aspects of connected vehicles, their applications will be proliferated in the upcoming years such as driverless cars and automated fleets. This may result in increased disputes as a result of their incidents. Therefore, we believe that there is a vast opportunity to pursue additional research with respect to vehicular forensics in general and our framework in particular. We list them below: 

\begin{itemize}
\item There will be a need to analyze the storage and communication overhead of B4F framework by implementing it using an OBU SDK. 
\item A punishment/incentive/avoidance mechanism should be investigated to prevent the members to become malicious actors. 
In this regard, a detection mechanism should be developed to discover malicious participants.
\item The B4F provides a lightweight solution by just keeping  hash values. While this ensures integrity and immutability of forensic data, the availability of this data depends on the individual storage and shared counterparts. There is no mechanism for ensuring availability of critical forensic data on blockchain. Therefore, this warrants further research. 
\item Due to increased availability of data and blockchain technologies in various domains for forensic purposes, researcher would need to consider forensic-by-design principle when proposing new systems and mechanisms. 
\item Regulations for enforcing the participation of various entities to forensic blockchains and development of policies to use such data in criminal cases are potential research issues.
 
\end{itemize}


\section{Conclusion}


In this paper, we proposed constructing a blockchain infrastructure to provide comprehensive forensic services for accident investigations. To address the issues regarding the overhead of storage and membership management of blockchain, we proposed using VPKI in permissioned blockchain and fragmented ledger which enables storage of hashed data in the shared ledger while the details are stored in fragmented ledgers as non-hashed data. In addition, the use of pseudonyms for identities help preserve privacy of the users.




\bibliographystyle{IEEEtran}
\bibliography{main}

\end{document}